\documentclass[12pt]{amsart}
\usepackage[mathscr]{eucal}
\usepackage{amsmath,amsfonts}

\parskip=\smallskipamount
                                                                                
\newtheorem{prop}{Proposition}
\newtheorem{thm}{Theorem}

\newtheorem{remark}{Remark}

\makeatletter
\@addtoreset{equation}{section}
\makeatother

\newdimen\expt
\def\boxit#1{\setbox0\hbox{$\displaystyle{#1}$}
      \hbox{\lower.4\expt
 \hbox{\lower3\expt\hbox{\lower\dp0
      \hbox{\vbox{\hrule height.4\expt
 \hbox{\vrule width.4\expt\hskip3\expt
      \vbox{\vskip3\expt\box0\vskip2\expt}%
 \hskip3\expt\vrule width.4\expt}\hrule height.4\expt}}}}}}
\begin{document}
\pagestyle{plain}

\bigskip

\title {Dilation Theoretic Parametrizations of Positive Matrices
With Applications to Quantum Information}

\author{M. C. Tseng \and V. Ramakrishna\\
Department of Mathematical Sciences\\
University of Texas at Dallas} 

\maketitle

\begin{abstract}
In this note, dedicated to the memory of Professor Tiberiu Constantinescu,
we discuss the applications of two parametrizations of positive matrices
to issues in quantum information theory. The first, which we propose
be dubbed the Schur-Constantinescu parametrization, is used in a twin fashion
to construct examples of separable states in arbitrary dimensions.
The second, called the Jacobi parametrization, is used to describe
quantum states in dimension two, as an alternative to the Bloch sphere 
representation. 
\end{abstract}

\section{Introduction}

Tiberiu Constantinescu, mathematician, mentor and colleague par excellence,
was interested in the rapidly developing field of quantum information
theory during the last few years of his life, \cite{tiviI,tiviII}.
In particular in \cite{tiviII}, several useful applications of the
Schur parameters to quantum information theory were presented.
Thus, for instance, the fact that the Schur parametrization automatically
yields the Cholesky factorization of the positive matrix in question implies
that a Kraus operator representation for quantum channels can be directly
found without any need for spectral factorization. In particular, viewing
a quantum state as a channel with one-dimensional input space, one can 
thereby find an ensemble representation for quantum states, which is
different from that provided by the state's spectral decomposition.
Similarly, the fact that the determinant of a positive matrix can be 
directly found from its Schur parameters is useful in computing some
entropic quantities. The Schur parameters were also shown to yield a
{\it simply verified} necessary and sufficient condition for purity (i.e., for
the quantum state to have rank one). This is particularly interesting
since there have been several attempts to extend the Bloch sphere 
condition for pure states in dimension two, \cite{Werner,nie,preskill}, 
to higher
dimensions. Whilst necessary and sufficient conditions can be stated, these
conditions are typically difficult to analyse. In this note, with a view
towards further popularizing this parametrization, we show how it can be
used to construct examples of separable states in a non ad-hoc fashion.

\begin{remark}
{\rm Whilst several researchers have made impressive contributions to
the field of Schur analysis (see \cite{Ba,Co} and the extensive bibliography
therein), it is our humble opinion that it was Tiberiu
Constantinescu that brought forth the versatility of this field to the
maximum. Thus, we propose that these parameters be dubbed the 
Schur-Constantinescu parameters (SC, for short), at least for the
purposes of this note.}
\end{remark}

In his most recent work, Tibi and his student developed a second
parametrization of positive matrices called the Jacobi parametrization,
\cite{CN,Es}, which loosely speaking is to positive Hankel matrices
what the SC parameterization is to positive Toeplitz matrices. This
parametrization also yields the Cholesky factorization and the determinant
of the positive matrix being parametrized, and thus would yield applications
similar to those in \cite{tiviII}. However, unlike the SC parametrization,
the Jacobi parametrization lacks an inheritance principle (see \cite{CN}),
and thus its utility, pending further investigation, seems limited in
comparison to the SC parametrization. Instead of reporting 
results on Kraus representations etc., analogous to those in \cite{tiviII},
we will restrict ourselves to showing how qubit states can be described
via the Jacobi parametrization. At this point we do not have a simple criterion
for purity in higher dimensions analogous to that produced by the
SC parametrization (and this is, in part, due to the lack of this
inheritance principle), and thus we will restrict ourselves to qubit states
in this note.

Partly with a view towards convincing researchers in quantum information
theory of the utility of the SC and Jacobi parametrization, and partly
out of deference to Tibi's abiding dedication to didaction, we will
present a ``user friendly" informal derivation of these two parametrizations.
We hope that this will stimulate novices in the field of Schur analysis
(as indeed, both authors of this note themselves are) to peruse Tibi's
innumerable contributions to this field.

The balance of this note is organized as follows. In the next section
the SC parametrization is presented. The next section contains our
results on separability. The fourth section presents the Jacobi
parametrization. The fifth describes the Jacobi picture of qubit states.
The final section offers conclusions.   

\section{SC parametrization}

We only provide the essential results required, together with
informal proofs where possible -  
the definitive and comprehensive treatment 
of SC parameters is given in \cite{Co}.   
In the following discussion, 
${\mathcal H}$, with subcripts, will denote a Hilbert space. 
${\mathcal L}({\mathcal H}_i , {\mathcal H}_j)$ is the 
space of bounded operators from 
${\mathcal H}_i$ to 
${\mathcal H} _j$ and ${\mathcal L}({\mathcal H} , {\mathcal H}) = 
{\mathcal L}({\mathcal H})$.\\ 

\noindent
Let
\[
A = 
\left[ 
\begin{array}{cc}
A_{11} & A_{12} \\
A_{12}^* & A_{22}
\end{array}
\right]
\in {\mathcal L}( {\mathcal H}_1 \oplus {\mathcal H} _2 )
\] 

\noindent
be a positive semidefinite operator matrix whose entries are bounded operators, that is 

\[
\langle \left[ \begin{array}{c} h_1 \\ h_2\end{array} \right], A  \left[\begin{array}{c} h_1 \\ h_2\end{array}\right] \rangle 
_{{\mathcal H}_1 \oplus {\mathcal H}_2} 
\geq 0
\] 

\noindent
for all $h_1 \in {\mathcal H} _1$ and $h_2 \in {\mathcal H} _2$.
Clearly the diagonal blocks themselves are then positive.
We will show that there exists an unique contraction 
$\Gamma \in {\mathcal L}({\mathcal H}_2 , {\mathcal H}_1)$ such that $A_{12} = A_{11} ^{\frac{1}{2}} \Gamma A_{22} ^{\frac{1}{2}}$. To this end,
assume for the moment that both $A_{11}$ and $A_{22}$ are invertible. 
Then a {\sl Frobenius-Schur identity} holds:

\[
A = 
\left[ 
\begin{array}{cc}
A_{11} & A_{12} \\
A_{12}^* & A_{22}
\end{array}
\right]
=
\left[ 
\begin{array}{cc}
I & 0 \\
A_{12}^* A_{11}^{-1} & I
\end{array}
\right]
\cdot
\left[ 
\begin{array}{cc}
A_{11} & 0\\
0 & A_{22} - A_{12} ^* A_{11}^{-1} A_{12}
\end{array}
\right]
\cdot
\left[ 
\begin{array}{cc}
I & A_{11} ^{-1} A_{12} \\
0 & I
\end{array}
\right].
\]

\noindent We dedude that $A$ is positive iff its Schur complement 

\[
A_{22} - A_{12} ^* A_{11}^{-1} A_{12} = 
A_{22} ^{\frac{1}{2}} (I - A_{22} ^{-\frac{1}{2}} A_{12} ^* A_{11}^{-1} A_{12} A_{22} ^{-\frac{1}{2}} ) A_{22} ^{\frac{1}{2}} 
\]

\noindent is positive, which implies

\[
I - A_{22} ^{-\frac{1}{2}} A_{12} ^* A_{11}^{-1} A_{12} A_{22} ^{-\frac{1}{2}}  \geq 0
\]

\noindent  The above can then be written as $(I -\Gamma^{*}\Gamma) \geq 0$.
In other words, the suitable contraction is 
$\Gamma = A_{11}^{-\frac{1}{2}} A_{12} A_{22} ^{-\frac{1}{2}}$.\\

\noindent
For the more general
case that $A_{11}$ and $A_{22}$ are not invertible, consider the sequences $\{ \alpha_n = A_{11} + \frac{1}{n} \}$ and 
$\{\beta _n = A_{22} + \frac{1}{n} \}$. By the spectral mapping theorem for self adjoint operators, $\alpha _n$ and $\beta _n$ are 
invertible for all $n$. Therefore there exist contractions $\{ \Gamma _n\}$ with 
$A_{12} = \alpha _n ^{\frac{1}{2}}\Gamma _n \beta _n ^{\frac{1}{2}}$. Since the unit ball in 
${\mathcal L}({\mathcal H}_2, {\mathcal H}_1)$ is compact in the weak operator topology, $\Gamma _n$ converges to some contraction $\Gamma$ weakly.
We can compute directly, for all $h_1 \in {\mathcal H} _1$ and 
$h_2 \in {\mathcal H} _2$,

\[
\langle h_1, A_{11} ^{\frac{1}{2}} 
\Gamma A_{22} ^{\frac{1}{2}} h_2 \rangle _{ {\mathcal H}_1 }
= \lim_n \langle h_1, A_{11} ^{\frac{1}{2}} 
\Gamma_n A_{22} ^{\frac{1}{2}} h_2 \rangle _{ {\mathcal H}_1 }
= \lim_n \langle h_1, \alpha_n ^{\frac{1}{2}} 
\Gamma_n \beta _n ^{\frac{1}{2}} h_2 \rangle _{ {\mathcal H}_1 }
= \langle h_1, A_{12} h_2 \rangle _{ {\mathcal H}_1 }.
\]

\noindent To summarize the above calculations:
\begin{thm}The operator matrix 
\[
A = \left[ \begin{array}{cc} A_{11} & A_{12} \\ A_{12}^* & A_{22} \end{array} \right]
\]
\noindent with $A_{ii}\geq 0$
\noindent
is positive iff there exists a contraction $\Gamma$ such that $A_{11} ^{\frac{1}{2}} \Gamma A_{22} ^{\frac{1}{2}} = A^{12}$. 
\end{thm} 

\noindent In the argument given, the unique positive square roots of $A_{11}$ and $A_{22}$ were used. In fact 
$A_{11}^{\frac{1}{2}}$ and $A_{22}^{\frac{1}{2}}$ can be replaced by any operator $L_1$ and $L_2$ satisfying $A_{11} = L_1 ^* L_1$ and 
$A_{22} = L_2 ^* L_2$. 

\noindent We next explain 
the crucial link between the these contractions and the
Cholesky factorization.
 For a contraction $\Gamma$, we introduce its {\sl defect operator} $D_{\Gamma} = (I - \Gamma ^* \Gamma)^{\frac{1}{2}}$. Since
$\Gamma (I - \Gamma ^* \Gamma) = (I - \Gamma \Gamma^*) T$, it follows that 
$\Gamma D_{\Gamma} = D_{\Gamma^*} \Gamma$. 
The Frobenius-Schur identity 
leads to a Cholesky factorization of $2 \times 2$ operator matrices:

\[
A = 
\left[ 
\begin{array}{cc}
L_1 ^* L_1 & L_1^* \Gamma L_2 \\
L_2^* \Gamma^* L_1 & L_2^* L_2
\end{array}
\right]
=
\left[ 
\begin{array}{cc}
L_1 ^* & 0 \\
L_2^* \Gamma^* & L_2^* D_{\Gamma}
\end{array}
\right]
\left[ 
\begin{array}{cc}
L_1 & \Gamma L_2 \\
0   &  D_{\Gamma} L_2
\end{array}
\right].
\] 

With a view towards extending this to operator matrices of size larger
than $2\times 2$, we first present the following result:

\begin{thm} {\rm
An operator $T = \left[ T_1 \; T_2 \; \cdots \;T_n \right] \in 
{\mathcal L}(\oplus _i ^n {\mathcal H} _i, {\mathcal H})$ is a contraction
iff $T_1 = \Gamma _1$ is a contraction and there exist $\Gamma _i$ such that 
$T_i = D_{\Gamma_1^*} D_{\Gamma_2 ^*} \cdots D_{\Gamma_{k-1}}^* \Gamma_k$}
\end{thm}

\noindent The proof follows by induction and is omitted.

\noindent
Having a characterization of row 
contractions, we now examine the $3 \times 3$ case. 
From theorem 1, an arbitrary positive 
$A \in {\mathcal L} ( \oplus_{i = 1} ^3 {\mathcal H}_i )$ is of the form:

\[
A = 
\left[ 
\begin{array}{ccc}
L_1 ^* L_1               & L_1^* \Gamma_{12} L_2     & B\\
L_2^* \Gamma_{12} ^* L_1 & L_2^* L_2                 & L_2 ^* \Gamma_{23} L_3\\
B^*                      & L_3^* \Gamma_{23} ^* L_2  & L_3 ^* L_3
\end{array}
\right]
\]

\noindent
where the $\Gamma_{ij}$'s are contractions. 
We will show that $B$ can be parametrized by contractions. Again from 
theorem 1, 

\[
\left[ L_1^* \Gamma_{12} L_2 \;\; B \right] = L_1 ^* \cdot \Lambda \cdot 
\left[ 
\begin{array}{cc}
L_2 & \Gamma_{23} L_3 \\
0   &  D_{\Gamma_{23}} L_3
\end{array}
\right]
\]

\noindent 
where $\Lambda \in {\mathcal L} ({\mathcal H}_2 \oplus 
{\mathcal H}_3, {\mathcal H}_1)$ is a contraction. By theorem 2, we can assume 

\[
\Lambda = \left[ \Lambda_1 \; \; D_{\Lambda_1 ^*} \Lambda_2 \right]
\]

\noindent for contractions $\Lambda _i$. Evidently we can choose $\Lambda_1 = \Gamma_{12}$ and let $\Lambda_2 = \Gamma_{13}$ to obtain

\[
B = L_1 ^* (\Gamma_{12} \Gamma_{23} + D_{\Gamma_{12}^*} \Gamma_{13} D_{\Gamma_{23}}) L_3 ^*.
\]

\noindent This can be generalized to positive operator matrices of arbitrary size. Let 
$A = [ A_{ij} ]_{ij} \in {\mathcal L}(\oplus _{i=1} ^n {\mathcal H}_i)$ 
be positive. The SC parametrization of $A$ can be calculated recursively
as follows:\\

i) $\left[ \begin{array}{cc} A_{n-1, n-1} & A_{n-1, n}\\ A_{n, n-1} & A_{n, n} \end{array} \right]$ is positive and can be parametrized 
according to 

the $2 \times 2$ case.

ii) For $1 \leq k \leq n-2$, the SC parametrization of
$\left[ \begin{array}{cccc} A_{k, k} & A_{k, k+1} & \cdots & A_{k,n}\\ A_{k+1, k} & A_{k+1, k+1}  & \cdots & A_{k+1, n}\\ \vdots & \vdots        
& \ddots & \vdots\\ A_{n, k} & A_{n, k+1}  & \cdots & A_{n,n} \end{array} \right]$ 

is calculated by first considering 
$\left[ \begin{array}{ccc} A_{k+1, k+1}  & \cdots & A_{k+1, n}\\ \vdots & \ddots & \vdots \\ A_{n, k+1} & \cdots 
& A_{n,n} \end{array} \right] = L_{k+1, k+1}^* L_{k+1, k+1}$, 

where $L_{k+1}$ is the Cholesky factor calculated in the previous step. 

Then put $[ A_{k, k+1} \; \cdots \; A_{k,n} ] = A_{k, k} ^{\frac{1}{2}} R_k L_{k+1}$ with $R_k$ being the corresponding 

row contraction.\\

\noindent The contractions $\Gamma_{ij}, j > i$ are called the
Schur-Constantinescu (SC) parameters of the matrix $A$. If the entries
of $A$ are scalars these are numbers in the unit disc. For a
positive matrix with scalar entries we adopt 
the convention that $\Gamma_{ij} = 0$ whenever
$A_{ii}A_{jj} = 0$ (see \cite{tiviII}).
Whilst applying the above iterative procedure leads to long 
expressions for the $\Gamma_{ij}$, these formulae can be easily 
represented via a transmission line diagram (this is also
closely related to the notion of Dyck paths, see \cite{Co,CN}), and thus
can be easily written down.  
The figure below illustrates this for the $4 \times 4$ case.  
This gives an easy way to compute the parameters. Each entry 
of the positve semidefinite $\{ A_{ij} \}$ corresponds 
to those paths in the diagram that startfrom $L_{jj}$ and end at ${L_{ii}}^*$. For example, each path from $L_{33}$ to ${L_{11}}^*$ 
describes to a summand in the parametrization of $A_{13}$.

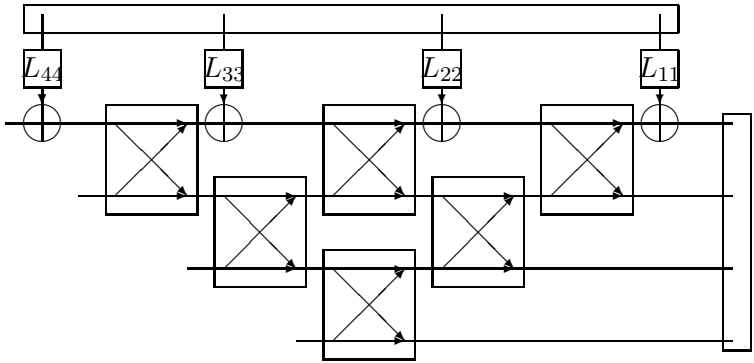
\begin{figure}[h]
\setlength{\unitlength}{3000sp}%
\begingroup\makeatletter\ifx\SetFigFont\undefined%
\gdef\SetFigFont#1#2#3#4#5{%
  \reset@font\fontsize{#1}{#2pt}%
  \fontfamily{#3}\fontseries{#4}\fontshape{#5}%
  \selectfont}%
\fi\endgroup%
\begin{picture}(6174,2949)(289,-2323)
{ \thinlines
\put(601,-361){\circle{300}}
}%
{ \put(2101,-361){\circle{300}}
}%
{ \put(3901,-361){\circle{300}}
}%
{ \put(5701,-361){\circle{300}}
}%
{ \put(301,-361){\line( 1, 0){900}}
}%
{ \put(601,-211){\line( 0,-1){300}}
}%
{ \put(1201,-361){\vector( 1,-1){600}}
}%
{ \put(1201,-961){\vector( 1, 1){600}}
}%
{ \put(1126,-1111){\framebox(750,900){}}
}%
{ \put(901,-961){\line( 1, 0){1200}}
}%
{ \put(1201,-361){\line( 1, 0){1200}}
}%
{ \put(2926,-1111){\framebox(750,900){}}
}%
{ \put(2026,-1711){\framebox(750,900){}}
}%
{ \put(3826,-1711){\framebox(750,900){}}
}%
{ \put(4726,-1111){\framebox(750,900){}}
}%
{ \put(2926,-2311){\framebox(750,900){}}
}%
{ \put(2101,-961){\line( 1, 0){1200}}
}%
{ \put(3301,-961){\line( 1, 0){1200}}
}%
{ \put(4501,-961){\line( 1, 0){1200}}
}%
{ \put(2401,-361){\line( 1, 0){1200}}
}%
{ \put(3601,-361){\line( 1, 0){1200}}
}%
{ \put(4801,-361){\line( 1, 0){1200}}
}%
{ \put(1201,-361){\vector( 1, 0){600}}
}%
{ \put(1201,-961){\vector( 1, 0){600}}
}%
{ \put(3001,-361){\vector( 1, 0){600}}
}%
{ \put(3001,-961){\vector( 1, 0){600}}
}%
{ \put(2101,-961){\vector( 1, 0){600}}
}%
{ \put(2101,-1561){\vector( 1, 0){600}}
}%
{ \put(3001,-1561){\vector( 1, 0){600}}
}%
{ \put(3001,-2161){\vector( 1, 0){600}}
}%
{ \put(3901,-961){\vector( 1, 0){600}}
}%
{ \put(3901,-1561){\vector( 1, 0){600}}
}%
{ \put(4801,-361){\vector( 1, 0){600}}
}%
{ \put(4801,-961){\vector( 1, 0){600}}
}%
{ \put(3001,-961){\vector( 1, 1){600}}
}%
{ \put(4801,-961){\vector( 1, 1){600}}
}%
{ \put(3001,-361){\vector( 1,-1){600}}
}%
{ \put(4801,-361){\vector( 1,-1){600}}
}%
{ \put(2101,-1561){\vector( 1, 1){600}}
}%
{ \put(3901,-1561){\vector( 1, 1){600}}
}%
{ \put(3001,-2161){\vector( 1, 1){600}}
}%
{ \put(3001,-1561){\vector( 1,-1){600}}
}%
{ \put(3901,-961){\vector( 1,-1){600}}
}%
{ \put(1801,-1561){\line( 1, 0){3000}}
}%
{ \put(2701,-2161){\line( 1, 0){1200}}
}%
{ \put(2101,-511){\line( 0, 1){300}}
}%
{ \put(3901,-511){\line( 0, 1){300}}
}%
{ \put(5701,-511){\line( 0, 1){300}}
}%
{ \put(451,-61){\framebox(300,300){\small $L_{44}$}}
}%
{ \put(3751,-61){\framebox(300,300){\small $L_{22}$}}
}%
{ \put(5551,-61){\framebox(300,300){\small $L_{11}$}}
}%
{ \put(451,389){\framebox(5400,225){}}
}%
{ \put(601,539){\line( 0,-1){300}}
}%
{ \put(3901,539){\line( 0,-1){300}}
}%
{ \put(5701,539){\line( 0,-1){300}}
}%
{ \put(601,-61){\vector( 0,-1){150}}
}%
{ \put(601,-61){\vector( 0,-1){150}}
}%
{ \put(1951,-61){\framebox(300,300){\small $L_{33}$}}
}%
{ \put(2101,539){\line( 0,-1){300}}
}%
{ \put(2101,-61){\vector( 0,-1){150}}
}%
{ \put(3901,-61){\vector( 0,-1){150}}
}%
{ \put(5701,-61){\vector( 0,-1){150}}
}%
{ \put(6001,-361){\line( 1, 0){300}}
}%
{ \put(5701,-961){\line( 1, 0){600}}
}%
{ \put(4801,-1561){\line( 1, 0){1500}}
}%
{ \put(3901,-2161){\line( 1, 0){2400}}
}%
{ \put(6226,-2236){\framebox(225,1950){}}
}%
{ \put(2101,-961){\vector( 1,-1){600}}
}%
\end{picture}

\caption{Cholesky factorization for $4\times 4$ positive matrices}
\end{figure}

\begin{remark}
{\rm The terminology scalar matrix will be used in the sequel to indicate that
a certain matrix is being thought of as having entries which are
$1\times 1$, in order to distinguish it from the general operator
matrix case. It should not be confused with 
a matrix of the form $\lambda I, \lambda \in C$.}
\end{remark}

\section{SC parameters and Separability}
\noindent We assume that the notions of density matrix, pure states
(i.e, rank one density matrices), positive preserving, complete positivity,
Kraus representations and Choi matrix are known to the reader
(see \cite{Werner,nie}).
When there is no chance of confusion, we will refer to a positive
preserving map, $\Phi$ simply as {\it positive}.

\noindent A quantum state on a tensor product of
two Hilbert spaces is said to be {\it separable} if it can be expressed
as a convex combination of product states. Characterizing separability   
is an important problem, since its violation, viz., the phenomenon of
quantum entanglement plays an essential role in quantum information
protocols. For $2\otimes 2$ and $2\otimes 3$ states there is a simple
necessary and sufficient condition for separability, the so-called
Peres-Horodecki PPT condition, \cite{Pe,Hor}. For other dimensions the only
known necessary and sufficient conditions are difficult to work
with in practice, \cite{Werner}.
The following result by the 
Horodecki's showing the existence 
of so-called entanglement witnesses is well known \cite{Hor}:

\noindent
\begin{thm} A mixed state $\sigma$ is separable iff $Tr(A \sigma) \geq 0$, for any bounded operator $A$ satisfying 
$Tr(A \cdot P \otimes Q) \geq 0$, for all product pure state $P \otimes Q$.\\
\end{thm}

\noindent
The above result was used by the Horodecki's to show a bijection between entanglement witnesses and non-CP maps.
This yields the following result which will be used to construct examples
of separable states
\noindent
\begin{thm}If a mixed state $\sigma \in {\mathcal L}( {\mathcal H}_A ) 
\otimes {\mathcal L}( {\mathcal H}_B )$ is such that
for every positive map $\Phi$ from ${\mathcal L}({\mathcal H}_B)$ to 
${\mathcal L}({\mathcal H}_A)$, the operator $(I_A \otimes \Phi)(\sigma)$ 
is positive, then $\sigma$ is separable.\end{thm}

From the discussion above, we see that if one has a certainly 
family ${\mathcal S}$ of states where any positive map restricted to 
${\mathcal S}$
is a completely positive map, 
then $\{S\}$ must consist of only separable states, due to lack of entanglement witnesses. We now obtain
such families via the factorization of quadratic forms with matrix coefficients. Namely, we will apply the following result due to 
Calderon \cite{Ca}:

\noindent
\begin{thm} Let $F$ be a quadratic form $F(s,t) = \alpha_1 s^2 + \alpha_2 st + \alpha_3 t^2$ whose coefficients 
$\alpha_i$ lie in $C^{m \times m}$ and the indeterminates $s$ and $t$ are real. If $F$ is positive for all $s$ and $t$ then
$F(s,t) = ( \beta_1 s + \beta_2 t )^* ( \beta_1 s + \beta_2 t )$, with $\beta_i \in C^{k \times m}$.\end{thm}

\noindent
Choi applied the above result and obtained that if $\Phi: C^{2 \times 2} \rightarrow C^{m \times m}$ is positive, then there 
exists $V_i \in C^{2 \times m}$ s.t. 
$\Phi(A) = \sum _i V_i ^* A V_i$ 
for all symmetric $A \in C^{2 \times 2}$. In other words,  
\cite{Ch} effectively showed that:

\begin{prop}Let ${\mathcal S}$ denote the set of $2 \times 2$ symmetric matrices. Any mixed state $\sigma \in C^{m \times m} \otimes S$ is 
separable \end{prop}

\noindent
\begin{bf}Proof\end{bf}: Let $\sigma$ be a mixed state of the form

\[
\sigma =
\left[
\begin{array}{ccc}
\sigma_{11} & \cdots & \sigma_{1m}\\
\vdots & \ddots & \vdots\\
\sigma_{m1} & \cdots & \sigma_{mm}
\end{array}
\right]
\]

\noindent, where each $\sigma _{ij}$ is $2 \times 2$ symmetric. For any positive map $\Phi : C^{2 \times 2} \rightarrow C^{m \times m}$, we
have
 
\[
(I_m \otimes \Phi)(\sigma) = 
\sum_k
\left[
\begin{array}{ccc}
V_k ^* \sigma_{11} V_k & \cdots & V_k ^* \sigma_{1m} V_k\\
\vdots                 & \ddots & \vdots\\
V_k ^* \sigma_{m1} V_k & \cdots & V_k ^* \sigma_{mm} V_k
\end{array}
\right]
= \sum_k (I \otimes V_k)^* \sigma (I \otimes V_k) \geq 0.  
\diamondsuit
\]

\begin{remark}
{\rm In particular, if a $2m\times 2m$ positive matrix is Hankel, when
viewed as a scalar matrix, then it is separable as a $m\otimes 2$ state.}
\end{remark}

\noindent
Using the SC parameters, 
this can be generalized to higher 
dimensions in a systematic way, although we will still be 
restricted to positive 
matrices that can be parametrized by two real parameters. To illustrate
this, consider the following $3 \times 3$ case.

\begin{prop} Let $A \in C^{3 \times 3}$ be of the form
\[
A =
\left[
\begin{array}{ccc}
a & a & b\\
a & a & b\\
b & b & c
\end{array}
\right]
\]

\noindent
For all positive $\Phi: C^{3 \times 3} \rightarrow C^{m \times m}$, where $m$ is arbitrary, there exists $V_i \in C^{3 \times m}$
s.t. $\Phi(A) = \sum _i V_i ^* A V_i$.
\end{prop}

\noindent
proof: By positivity of $\Phi$, we have 

\[
\Phi 
(
\left[
\begin{array}{ccc}
s^2 & s^2 & st\\
s^2 & s^2 & st\\
st  & st  & t^2
\end{array}
\right]
) \geq 0.
\]

\noindent
The positivity of the matrix in the argument of 
$\Phi$ can be directly checked using the SC parametrization. 
One can simply pick all the contractions in the SC parametrization of
positive scalar matrices to be 1. 
This means the following quadratic form:

\[
F(s,t) = s^2 \cdot  \; \Phi (E_{11} \; + \; E_{12} + E_{21} + E_{22}) \; + \; st \cdot \; \Phi(E_{13} + E_{23} + E_{31} + E{32}) \; + \;
t^2 \cdot \; \Phi(E_{33}) 
\geq 0
\]

\noindent ,for all real $s$ and $t$. By Calderon's result, we have $\Phi (E_{11} + E_{12} + E_{21} + E_{22}) = C^* C$, 
$\Phi(E_{33}) = D^* D$, and $\Phi(E_{13} + E_{23} + E_{31} + E_{32}) = C^* D + D^* C$, where $C, D \in C^{k \times m}$ for some 
integer $k$. Let $\Phi ' : C^{3 \times 3} \rightarrow C^{m \times m}$ be defined by its Choi matrix:

\[
M_{\Phi '} = ( \Phi ' (E_{ij}) )_{ij} =
\left[
\begin{array}{ccc}
\frac{1}{4}C^*C   & \frac{1}{4}C^*C    & \frac{1}{2} C^* D\\
\frac{1}{4}C^*C   & \frac{1}{4}C^*C    & \frac{1}{2} C^* D\\
\frac{1}{2} D^* C & \frac{1}{2} D^* C  & D^* D
\end{array}
\right].
\]

\noindent Again from the SC parametrization, 
the above matrix is manifestly positive. So $\Phi '$ is completely positive. 
If $A$ is of the 
specified form, then

\[
\begin{array}{l} 
\Phi(A) = a \cdot  \Phi (E_{11} + E_{12} + E_{21} + E_{22}) + b \cdot \Phi(E_{13} + E_{23} + E_{31} + E{32}) + c \cdot \Phi(E_{33})\\
\;\\
= a \cdot C^*C + b \cdot (C^* D + D^* C) + c \cdot D^*D\\
\; \\
=a \cdot  \Phi' (E_{11} + E_{12} + E_{21} + E_{22}) + b \cdot \Phi' (E_{13} + E_{23} + E_{31} + E{32}) + c \cdot \Phi '(E_{33})
\; \\
= \Phi ' (A).
\end{array}
\]

\noindent
Now take $\{V_i\}$ to be any set of Kraus operators of $\Phi'$ and this completes the proof. $\diamondsuit$\\

\noindent
This translates to:

\begin{prop}
Let ${\mathcal S}_1 \subset C^{3 \times 3}$ denote the family of matrices of the same form as $A$ in the above proposition, then any mixed state
$\rho \in C^{m \times m} \otimes S_1$, where $m$ is arbitrary, is separable.
\end{prop}

\noindent It is evident from the SC parameters that ${\mathcal S}_1$ can be replaced by:

\[
{\mathcal S} _2 = \{ A \in C^{3 \times 3} | \;
A =
\left[
\begin{array}{ccc}
a & c & a\\
c & b & c\\
a & c & a
\end{array}
\right]
\}
\]

or

\[
{\mathcal S}_3 = \{ A \in C^{3 \times 3} | \;
A =
\left[
\begin{array}{ccc}
a & c & c\\
c & b & b\\
c & b & b
\end{array}
\right]
\}.
\]

The general procedure using the above approach is as follows. Let $n$ be the size of each block. In the
SC parametrization of positive scalar matrices, 
let all contractions be 1. Choose disjoint sets of indices $I_s$ and $I_t$ such that
$I_s \cup I_t = \{1,..., n\}$. Denote the cardinality of $I_s$ and $I_t$ by $n_s$ and $n_t$ respectively. Consider the $n$ diagonal
elements from $A = [a_{ij}] \in {\mathbf C}^{n \times n}$. SC 
parametrize $a_{ii}$ by $s^2$ if
$i \in I_s$ and $a_{ii}$ by $t^2$ if $i \in I_t$. For any positive map $\Phi$, $\Phi (A) \geq 0$. Linearity of $\Phi$ gives a
positive quadratic form with coefficients in ${\mathbf C}^{n \times n}$:
\[
\Phi(A) = s^2 (\sum_{i, j \in I_s} \Phi(E_{ij})) + st (\sum_{i \in I_s, i \in I_t} \Phi(E_{ij}) + \Phi(E_{ji})) +
t^2 (\sum_{i, j \in I_t} \Phi(E_{ij})) \geq 0
\]
                                                                                
\noindent for all $s$ and $t$. Calderon's result gives
                                                                                
\[
\sum_{i, j \in I_s} \Phi(E_{ij}) = C^*C, \sum_{i \in I_s, i \in I_t} \Phi(E_{ij}) + \Phi(E_{ji}) = (C^*D + D^*C),
\sum_{i, j \in I_t} \Phi(E_{ij}) = D^*D
\]
                                                                                
\noindent
for some $C$ and $D$. Define a linear map $\Psi$ by
                                                                                
\[
\Psi(E_{ij}) = \frac{1}{n_s ^2} C^* C, \; \forall i, j \in I_s ,
\]
\[
\Psi(E_{ij}) = \frac{1}{n_t ^2} D^* D, \; \forall i, j \in I_t ,
\]
                                                                                
\[
\Psi(E_{ij}) = \frac{1}{n_s n_t} C^* D, \; \forall i \in I_s, j \in I_t ,
\]
                                                                                
\noindent and

\[
\Psi(E_{ij}) = \frac{1}{n_s n_t} D^* C, \; \forall i \in I_t, j \in I_s.
\]
                                                                                
\noindent Inspecting the Choi matrix of $\Psi$, $( \Psi(E_{ij}))_{ij}$ shows that $\Psi$ is CP. Let $A'$ be obtained from $A$ by replacing
$s^2$, $t^2$, and $st$ by arbitrary complex numbers $a$, $b$, and $c$ respectively. Then
\[
\Phi(A') = a (\sum_{i, j \in I_s} \Phi(E_{ij})) + c (\sum_{i \in I_s, i \in I_t} \Phi(E_{ij}) + \Phi(E_{ji})) +
b (\sum_{i, j \in I_t} \Phi(E_{ij}))
\]
                                                                                
\[
= a \cdot C^*C + c \cdot (C^*D + D^*C) + b \cdot D^*D
\]
                                                                                
\[
= a \cdot n_s ^2 \frac{1}{n_s ^2} C^* C + c \cdot n_s n_t (\frac{1}{n_s n_t} C^* D +  \frac{1}{n_s n_t} D^* C) +
b \cdot n_t ^2 \frac{1}{n_t ^2} D^* D
\]
                                                                                
\[
= a (\sum_{i, j \in I_s} \Psi(E_{ij})) + c (\sum_{i \in I_s, i \in I_t} \Psi(E_{ij}) + \Psi(E_{ji})) +
b (\sum_{i, j \in I_t} \Psi(E_{ij}))
\]
\[
= \Psi(A').
\]
                                                                                
\noindent Thus any positive map 
$\Phi$ is CP when restriced 
to matrices of the form $A'$ specified above. Let ${\mathcal A}'$ be the
subspace of matrices of the form $A'$, and ${\mathcal S} = 
{\mathcal A}' \otimes {\mathbf C}^{k \times k}$ for arbitrary $k$.
The correspondence between 
positive but non-CP maps and entanglement witnesses implies the result below.
                                                                                
\begin{thm}Any state $\rho \in {\mathcal S}$ is separable.\end{thm}
                                                                                
In this approach to obtain separable states, the SC parametrization was used twice: to get a subfamily of positive matrices
that can be parametrized by two real parameters, and to ensure the proposed map $\Psi$ has a positive Choi matrix, therefore CP.
A natural question is whether this can be extended 
by varying the contractions in the parametrization of scalar positive matrices,
i.e., by picking contractions which are not all equal to $1$. 
To illustrate the issues at hand, consider a $3 \times 3$
case. Suppose one parametrize a subfamily of $3 \times 3$ positive matrices in the following way. Let
                                                                                
\[
A =
\left[
\begin{array}{ccc}
s^2               & s \alpha t &  s \beta  s\\
t {\bar \alpha} s & t^2        &  t \gamma s\\
s {\bar \beta}  s & s {\bar\gamma} t & s^2
\end{array} \right]
\geq 0
\]
                                                                                
\noindent
where $\alpha$, $\beta$, and $\gamma$ are specified by some chosen contractions $\Gamma_i$, $i = 1,2,3$, from the SC parameters. The
proposed map $\Psi$ would then have as its Choi matrix
                                                                                
\[
M_{\Psi} =
\left[
\begin{array}{ccc}
\frac{1}{2 + \beta + {\bar \beta}} C^*C  &  \frac{\alpha}{\alpha +{\bar \gamma}} C^*D &  \frac{\beta}{2 + \beta + {\bar \beta}} C^*C\\
\frac{{\bar \alpha}}{{\bar \alpha} + \gamma}D^*C & D^*D        &  \frac{\gamma}{{\bar \alpha} + \gamma} D^*C\\
\frac{{\bar \beta}}{2 + \beta + {\bar \beta}} C^*C & \frac{{\bar \gamma}}{\alpha + {\bar \gamma}} C^* D & \frac{1}{2 + \beta + {\bar \beta}} C^*C
\end{array} \right]
\]
                                                                                
\noindent where $C$ and $D$ are obtained the same way as before. To get $M_{\Psi} \geq 0$, we apply the operator version of
SC parametrization and find 
suitable contractions for the above operator matrix. 
This amounts to solving a system of nonlinear
inequations on unit disks. For instance, the $2 \times 2$ leading minor
\[
\left[
\begin{array}{cc}
\frac{1}{2 + \beta + {\bar \beta}} C^*C  &  \frac{\alpha}{\alpha +{\bar \gamma}} C^*D \\
\frac{{\bar \alpha}}{{\bar \alpha} + \gamma}D^*C & D^*D      \\
\end{array} \right]
\]
                                                                                
\noindent corresponds to the inequation
                                                                                
\[
| \frac{1}{(2 + \beta + {\bar \beta})^{\frac{1}{2}}} | \geq | \frac{\alpha}{\alpha +{\bar \gamma}} |
\]
                                                                                
\noindent in the variables $\Gamma_i$, $i = 1,2,3$. Determining the freedom in choosing $\Gamma_i$ amounts to solving the corresponding
system of inequations.

\noindent {\it Direct application of Schur parameters 
to obtain separable states}:
In the channel-state duality,
between states in ${\mathcal L}({\mathcal H}_A \otimes {\mathcal H}_B)$ and 
channels $\Phi: {\mathcal L}({\mathcal H}_A) 
\rightarrow {\mathcal L}({\mathcal H}_B)$,
any square-root factorization 
of the state gives a set 
of Kraus operators for the dual channel. In this duality the channel
$\Phi$'s dual (unnormalized) state is precisely the Choi matrix $M_{\Phi}$.
The following fact is easy to
verify:
 
\noindent {\it Let $\rho$ be a $n \times m$ bipartite mixed state.
$\rho$ is separable iff its dual channel has a Kraus representation
in which all non-zero Kraus operators have rank 1.}  
The difficulty with applying this condition is that Kraus representations
are not unique, and it is entirely conceivable that there are different
Kraus representations some of which have this property, while the
others do not. Hence, effectively
this condition is typically only sufficient.  
Now, the SC parametrization explicitly calculates the 
Cholesky factors of positive matrices. The utility of this to the present
context is that most of
the Kraus operators produced by the Cholesky factorization tend to be
sparse matrices (see \cite{tiviII} for the case of binary channels) and
{\it thus there is a greater chance of detecting a Kraus representation with
all Kraus operators having rank one}.   
We illustrate this via the case of $3\otimes 3$ states.

Let a $3 \otimes 3$ state $\rho \in {\mathbf C}^{9 \times 9}$ have the Cholesky factorizations calculated from the Schur parameters $\rho = F F^*$, where $F = (F_{ij})_{1 \leq i,j \leq 9}$. By inspection, we see that the corresponding Kraus operators all have rank
$1$ if all of the following hold:\\
                                                                                
\noindent
\begin{bf}1.\end{bf} One of $D_{\Gamma_{i6}}$, $i = 1,...,5$ is $0$, i.e. one of $| \Gamma_{i6} |$, $i = 1,...,5$, is $1$.\\
                                                                                
\noindent
\begin{bf}2.\end{bf} One of $D_{\Gamma_{i5}}$, $i = 1,...,4$ is $0$.\\
                                                                                
\noindent
\begin{bf}3.\end{bf} If $D_{\Gamma_{i6}} \neq 0$, for $i = 2,...,5$, then $D_{\Gamma_{16}} = 0$ and $|\Gamma_{16} = 1|$, in which
case we require $\Gamma_{15} = 0$.\\
                                                                                
\noindent
\begin{bf}3.\end{bf} We check via direct calculation if 
the Kraus operator corresponding to the fourth row of $F$ have rank $1$.\\
                                                                                
\noindent
\begin{bf}4.\end{bf} $D_{\Gamma_{23}} = 0$ or $D_{\Gamma_{13}} = 0$.\\
                                                                                
\noindent
\begin{bf}5.\end{bf} $| \Gamma_{12} | = 0$.\\
                                                                                
\noindent
\begin{bf}6.\end{bf} Finally, we check if the Kraus operator corresponding 
to the first row of $F$ have rank $1$.\\
                                                                                
\noindent As mentioned before, due to the non-uniqueness of Kraus
representations, these are effectively only sufficient conditions.
Indeed, if one were to apply a similar procedure to $2\otimes 2$ states, 
then there
are block-Toeplitz matrices, which are known to satisfy the Peres-Horodecki
PPT criterion (and would thus be separable), but which fail to satisfy the
analogous set of conditions.

\section{Jacobi parametrization}

While the SC parameters can 
be viewed as an extension of 
positive Toeplitz kernels, the Jacobi parameters generalize positve kernels
of Hankel type, \cite{CN,Es}. As with the SC case, 
we only give results relevant in the present context, together
with informal derivations. A detailed presentation can be found in
\cite{CN,Es}.\\

\noindent
For a positive Hankel matrix

\[
H = 
\left[
\begin{array}{ccccc}
s_0     & s_1     & s_2     & \cdots & s_n     \\
s_1     & s_2     & s_3     & \cdots & s_{n+1} \\
\vdots  & \vdots  & \ddots  & \vdots & \vdots  \\
s_n     & s_{n+1} & \cdots  & \cdots & s_{2n}
\end{array} \right]
\] 

\noindent
Then it is known that there exists a symmetric tri-diagonal 

\[
J = 
\left[
\begin{array}{ccccc}
b_0     & a_1     & 0       & \cdots &  0       \\
a_1     & b_1     & a_2     & \cdots &  \vdots  \\
0       & a_2     & \ddots  & \ddots &  \vdots  \\
\vdots  & \vdots  & \ddots  & \ddots &  a_n     \\
0       & 0       & \cdots  & a_n    &  b_n
\end{array} \right]
\]

\noindent such that (notice by symmetry and Hermiticity 
$a_i$'s are real while $b_i$'s are complex)

\[
s_n = \langle J^n e_0, e_0 \rangle, n > 0 
\]
and
\[
s_{0} = s_{0}\langle e_{0}, e_{0}\rangle
\] 

\noindent where $e_0 = \left[ \begin{array}{c} 1 \\ 0 \\ \vdots \\ 0 \end{array} \right]$ is the basis element in $(n+1)$-dimensional Hilbert
space. 
This yields the following Cholesky factorization of $H$:

\[
H = \left[ \begin{array}{c} \sqrt{s_{0}} e_0 ^* \\ (J e_0)^* \\ ( J^2 e_0 ) ^* \\ \vdots \\ (J^n e_0)^* \end{array} \right]
\left[ \begin{array}{ccccc} \sqrt{s_{0}}e_0 & J e_0 & J^2 e_0 & \cdots & J^n e_0 \end{array} \right] 
\]

\noindent Thus one  might 
speculate that arbitrary positive matrices 
might have similar structures. Specificly, given
an $(n+1) \times (n+1)$ positive matrix $(A_{ij})$, one may
speculate that there may exist tri-diagonal $(n+1) \times (n+1)$ matrices 
$\{ J_1, \cdots, J_n \}$ such that

\[ 
A_{ij} = \langle J_i \cdots J_1 e_0 \; , \; J_j \cdots J_1 e_0 \rangle.
\]

\noindent Such a parametrization is called a {\sl tri-diagonal model} of positive matrices. A tri-diagonal would lead to the 
Cholesky factorization

\[
A = \left[ \begin{array}{c} s_{0}e_0 ^* \\ (J_1 e_0)^* \\ (J_2 J_1 e_0)^* \\ \vdots \\ (J_n \cdots J_1 e_0)^* \end{array} \right] 
\left[ \begin{array}{ccccc} s_{0}e_0 & J_1 e_0 & J_2 J_1 e_0 & \cdots & J_n \cdots J_1 e_0 \end{array} \right] 
\]

\noindent It was shown in \cite{Es} that there exists positive matrices which have no tridiagonal model. Instead, a general
positive semi-definite can be described by what is called the {\sl near tri-diagonal model}. Namely, the suitable matrices are given
by

\[
J_1 = 
\left[
\begin{array}{ccccc}
b_0 & a_1 & 0    & 0       & \; \\
a_1 & b_1 & a_2  & 0       & \; \\
0   & a_2 & b_2  & a_3     & \; \\
0   & 0   & a_3  & b_3     & \; \\
\;  & \;  & \;   & \;      & \ddots 
\end{array} \right],
\]

\[
J_2 = 
\left[
\begin{array}{ccccc}
b_0 & c_{0,1} & 0    & 0       & \; \\
a_1 & b_1     & a_2  & 0       & \; \\
0   & a_2     & b_2  & a_3     & \; \\
0   & 0       & a_3  & b_3     & \; \\
\;  & \;      & \;   & \;      & \ddots 
\end{array} \right],
\]

\[
J_3 = 
\left[
\begin{array}{ccccc}
b_0 & c_{0,1} & c_{0,2}    & 0       & \; \\
a_1 & b_1     & c_{1,2}     & 0       & \; \\
0   & a_2     & b_2        & a_3     & \; \\
0   & 0       & a_3        & b_3     & \; \\
\;  & \;      & \;         & \;      & \ddots 
\end{array} \right],
\]

\[
\vdots
\]

\noindent
, where $a_i \geq 0$, $b_i$ and $c_{i,j}$'s are complex.\\

\noindent The theorem below says any $(n+1) \times (n+1)$ positive semidefinite matrix can then be described by 
$\{J_1, \cdots, J_n\}$ in the manner specified above. For the $2 \times 2$ case, this is to say that any
$A =  \left[ \begin{array}{cc} A_{00} & A_{01} \\ A_{10} & A_{11} \end{array} \right]\geq 0$ can be written in the form

\[
A = 
\left[
\begin{array}{cc}
s_0          & 0   \\
{\bar b_0}   & a_1 \\
\end{array} 
\right]
\left[
\begin{array}{cc}
s_0 & b_0 \\
0   & a_1 \\
\end{array} 
\right]
=
D_1 ^* D_1
=
\left[
\begin{array}{cc}
s_0 ^2          & s_0 b_0 \\
s_0 {\bar b_0}  & | b_0 | ^2 + a_1 ^2 \\
\end{array} 
\right]
\]

\noindent for $s_0 \geq 0$ ,which is obvious. To show a positive $3 \times 3$ 

\[
A=
\left[
\begin{array}{ccc}
A_{00} & A_{01} & A_{02}\\
A_{10} & A_{11} & A_{12}\\
A_{20} & A_{21} & A_{22}\\
\end{array}
\right]
\] 

\noindent
can be expressed via  near tri-diagonal model, i.e.

\[
A = 
\left[
\begin{array}{c}
( s_{0}e_0 ')^* \\ (J_1 e_0)^* \\ (J_2 J_1 e_0)^*
\end{array}
\right]
\left[
\begin{array}{ccc}
s_{0}e_0 ' & J_1 e_0 & J_2 J_1 e_0
\end{array}
\right]
\]

\[
=
\left[
\begin{array}{ccc}
s_0                                  & 0                        & 0\\
{\bar b_0}                          & a_1                      & 0\\
{\bar b_0}^2 + {\bar c_{0,1}} a_1   & a_1 ({\bar b_0 + b_1})   & a_1 a_2             
\end{array}
\right]
\left[
\begin{array}{ccc}
s_0 & b_0 & b_0 ^2 + c_{0,1} a_1\\
0   & a_1 & a_1 (b_0 + b_1)     \\
0   & 0   & a_1 a_2             
\end{array}
\right]
\]

\[
=
\left[
\begin{array}{cc}
D_1 ^*                                                                                 & 0 \\
\begin{array}{cc} {\bar b_0}^2 + {\bar c_{0,1}} a_1 & a_1 {\bar b_0 + b_1} \end{array} & a_1 a_2
\end{array}
\right]
\left[
\begin{array}{cc}
D_1 & \begin{array}{c} b_0 ^2 + c_{0,1} a_1 \\ a_1 (b_0 + b_1) \end{array} \\
0   & a_1 a_2
\end{array}
\right]
\]

\noindent The parameters 
$s_0$, $b_0$, and $a_1$ can be assumed to be known from the $2 \times 2$ case. 
Using the fact that there 
must exist a Cholesky factorization, there exist complex numbers
$x_{1}$ and $x_{2}$ such that 
$D_1  \left[ \begin{array}{c} x_1 \\ x_2 \end{array} \right] = 
\left[ \begin{array}{c} A_{02} \\  A_{12} \end{array} \right]$. 
Then $c_{0,1}$
and $b_1$ can be obtained by solving the equations $x_1 = b_0 ^2 + c_{0,1} a_1$ and $x_2 = a_1 (b_0 + b_1)$ respectively. Using the usual
argument for Cholesky factorizations one can calculate $a_2 \geq 0$ easily.\\

\noindent
Notice for the $3 \times 3$ case it is the parameter $c_{0,1}$ that allows the model to work. The general statement is:

\begin{thm}Given any $(n+1) \times (n+1)$ positive semidefinite matrix $A = ( A_{ij} ) _{0 \leq i,j, \leq n}$, there exists
a near tri-diagonal model $\{J_1, \cdots, J_n \}$. \end{thm}

\noindent (Notice this statement is slightly different from that of \cite{Es} in that 
$A_{0,0}$ is not assumed to be $1$ and positive definiteness is not required).\\
 
\noindent
\begin{bf}Proof\end{bf}: The proof uses exactly the same idea outline above. Assume the statement holds for the $n \times n$ case. 
We need to show

\[
A = 
\left[
\begin{array}{cc}
D_{n-1} & l_n   \\ 
0       & a_1 \cdots a_n \\
\end{array}
\right]
\]

\noindent where 

\[
l_n = J_n \cdots J_1 e_0
\]

\noindent and $a_n \geq 0$ is to be determined. Again using the fact that there exists a Cholesky factorization, one can find 
$\left[ \begin{array}{c} x_1 \\ \vdots \\ x_n \end{array} \right]$ such that 

\[
D_{n-1} \left[ \begin{array}{c} x_1 \\ \vdots \\ x_n \end{array} \right] =  
\left[ \begin{array}{c} A_{0,n} \\ \vdots \\ A_{n-1,n} \end{array} \right].
\]

\noindent The desired parameters can be obtained directly by solving the corresponding $n$ equations. $\diamondsuit$\\

\noindent As the combinatorial structure of SC 
parameters can be seen as given by the Dyck paths, the Jacobi parameters can 
be viewed via Lukasiewicz paths, \cite{CN,Es}.

\section{Jacobi parameters and $2 \times 2$ mixed states}
 As mentioned in the introduction the Jacobi parametrization also produces
the Cholesky factorization and yields a formula for the determinant
analogous to the case for the SC parametrization. We will not repeat 
applications of these features similar to that in \cite{tiviII}.
Instead, an elegant geometric picture for $2\times 2$ states which
is different from the so-called Bloch sphere emerges from the
Jacobi parametrization.
 
A $2 \times 2$ positive semidefinite matrix $A$ can be Jacobi parametrized by

\[
A = 
\left[
\begin{array}{cc}
s_0          & 0   \\
{\bar b_0}   & a_1 \\
\end{array} 
\right]
\left[
\begin{array}{cc}
s_0 & b_0 \\
0   & a_1 \\
\end{array} 
\right]
=
D_1 ^* D_1
=
\left[
\begin{array}{cc}
s_0 ^2          & s_0 b_0 \\
s_0 {\bar b_0}  & | b_0 | ^2 + a_1 ^2 \\
\end{array} 
\right]
\]

\noindent where $s_0 \geq 0$, $a_1 \geq 0$, and $b_0$ is complex. The trace $1$ requirement means 
$s_0 ^2 + a_1 ^2 + | b_0 | ^2 = 1$. This essentially says that the $2 \times 2$ mixed states lie in the following set in $R^4$:

\[
\{s_0 ^2 + a_1 ^2 + \alpha ^2 + \beta ^2  =1 | s_0 \geq 0, \; a_1 \geq 0\}.
\]

\noindent The pure states also have a 
nice description. 
We adopt the convention when $s_0 = 0$, $b_0 = 1$ (cf., the
convention for the SC parameters in \cite{tiviII}). 
The general condition 
for the state to be pure is the state is pure iff $a_1 = 0$. 
This gives a set in $R^3$:

\[
S_1 = \{s_0 ^2 + \alpha ^2 + \beta ^2  =1 | s_0 \geq 0 \}
\]

\noindent
,which is exactly the surface of the upper hemisphere in $R^3$. When $s_0 = 1$ and $|b_0| = 0$, this is the classical $| up \rangle$
state and it is at the pole of the hemisphere. When $s_0 = 0$ and $b_0 = 1$, this is the $| down \rangle$ state and it is the point
$(1,0)$ in the $xy$-plane.\\

\section{Conclusions}
This brief note presented applications
to the field of quantum information of two parametrizations
of positive operator matrices, the SC
and Jacobi parametrization, that Professor Tiberiu Constantinescu was
instrumental in developing.
These two applications are merely an indication of what is possible
with these parametrizations. In our humble opinion these two parametrizations
and the attendant characterizations of positive matrices are amongst the
most versatile such characterizations. Related to these parametrizations
are parametrizations of that other important class of matrices in quantum
theory, viz., unitary matrices, which this note did not dwell upon.
We hope that our presentation of these two parametrizations will stimulate
readers into pursuing the various fascinating aspects (especially the
combinatorial structures) of this subject in relation to quantum 
theory and its applications.

\end{document}